\documentstyle[epsfig,aps]{revtex}

\begin{document}
 
\title{Toy Model for Pion Production in Nucleon-Nucleon Collisions}
 
\author{
C. Hanhart$^a$, G.A. Miller$^a$, 
F. Myhrer$^b$, T. Sato$^{b,c}$, and U. van Kolck$^{d,e,f}$}

\address{
\small{$^a$Dept. of Physics, University of Washington,
Seattle, WA 98195-1560} \\
\small{$^b$Dept. of Physics and Astronomy, University of South Carolina,
Columbia, SC 29208} \\
\small{$^c$Dept. of Physics, Osaka University,
Toyonaka 560-0043, Osaka, Japan} \\
\small{$^d$Dept. of Physics, University of Arizona,
Tucson, AZ 85721}\\
\small{$^e$RIKEN-BNL Research Center, Brookhaven National Laboratory,
Upton, NY 11973}\\
\small{$^f$Kellogg Radiation Laboratory 106-38,
California Institute of Technology,
Pasadena, CA 91125} }

\maketitle 

\vskip .5cm \noindent

\begin{abstract}
We develop a toy model of pion production in nucleon-nucleon collisions
that reproduces some of the features of the chiral Lagrangian
calculations.
We calculate the production amplitude 
and examine some common approximations.
\end{abstract}

\vskip .5cm \noindent

\hfill{NT@UW--00--24}

\hfill{RBRC--142}

\hfill{KRL MAP--273}

\section{Introduction} 

Interest in studies of pion production in nucleon-nucleon collisions at
energies near threshold 
has been re-vitalized by the appearance of excellent high quality
data \cite{meyer}.
The fact that  low- and medium-energy strong interactions are controlled
by chiral symmetry led to           an early  hope that
 chiral effective theories could be used to
analyze these processes and achieve a fundamental understanding
of the production process. 
Indeed, 
 there are now tree level calculations 
\cite{cohen,tree,sato,hanhart98,vkmr,rocha00}
 and even 
loop calculations \cite{loops,dmitra99,ando00} available in the literature
\cite{chrisreview}.
The early 
excitement was quickly
abated by the realization that proper evaluation involves surmounting several
severe  difficulties, which are caused by the high momentum transfer nature of
this threshold process.  The initial relative
momentum between the two nucleons must be at least
$p_i=\sqrt{m_\pi M_N}$. This means
that  the 
chiral expansion
is in terms of powers  of $\sqrt{m_\pi/M_N}$ instead of
$m_\pi/M_N$ \cite{cohen,pwave},which complicates carrying out the expansion
and verifying its convergence. 
However, issues of convergence 
are not the focus
of the present  work. Instead,     
we address some technical questions which arise during 
the evaluation of the relevant matrix elements.

It is worthwhile 
to discuss some general features of the pion production
process 
before describing our        
 specific technical issues. 
Pion production occurs when the mutual interactions
between two nucleons cause a real
pion to be emitted. The leading term is one in which the initial and
 final state two-nucleon ($NN$) scattering allow a pion to be emitted by a
 single nucleon emission. The next tree level contribution occurs when 
a virtual pion of four-momentum $q$ produced by one nucleon is 
knocked on to its mass shell by an
interaction with  the second nucleon. This 
is the so-called re-scattering diagram. This process
 typically occurs accompanied by low-momentum-transfer initial and/or
 final state interactions.
  The
evaluation of these diagrams, including the case  
when        the 
pion exchanged between the two nucleons may be on shell, is our 
focus. Our strategy will be             
 to introduce a toy model, which is simple enough to allow the exact
evaluation of certain amplitudes. Then we may assess various          
approximations by comparing the resulting amplitudes to the exact results.

In general, one could obtain the necessary transition matrix elements by
evaluating the relevant  Feynman diagrams. However, the 
                 initial and final state interactions 
      are accurately treated using an appropriate $NN$ potential within a
three-dimensional Schr\"{o}dinger equation formulation. Thus one needs to
obtain a three-dimensional formulation from the more general Feynman procedure.
This has been done in an {\it ad hoc} manner in 
Refs. \cite{cohen,tree,vkmr,rocha00,loops}:
one guesses the energy dependence
of the virtual pion-nucleon ($\pi N$) interaction, 
and uses a Klein-Gordon propagator for          the pion propagator.   
However, there is a general method to 
derive a three-dimensional theory 
which is equivalent to the Feynman diagram approach, namely the
method of considering all the time-ordered diagrams ---the
use of time-ordered perturbation theory (TOPT). 
In this formulation, one finds only
$NN$, $\pi NN$ and $\pi \pi NN$ propagators 
in the tree-level re-scattering diagrams. 
The Feynman Klein-Gordon pion propagator does not appear explicitly.
Thus our first focus  is the appropriate propagator. 
In particular, we will
compare different prescriptions used in the literature with the exact
result derived in the toy model. 

Another issue to be addressed is that of the proper choice of
the energy variable $q_0$ of the exchanged pion.
The value of
$q_0$ is critical because the chiral $\pi N$ interaction includes  
seagull vertices involving $\partial_0 \pi$, such as
the isovector Weinberg-Tomozawa interaction 
$N^{\dagger}\tau N \cdot (\pi \times \partial_0 \pi)$
and the isoscalar
$N^{\dagger}N (\partial_0 \pi)^2$.
In case of the isoscalar re-scattering, which is most relevant
for threshold $\pi^0$ production, 
this seagull term is $\propto q_0 m_\pi$.
Its
actual size
is crucial: for on-shell 
$\pi N$ scattering at threshold ($q_0=m_\pi$) there is an almost complete
cancelation of different, individually large terms leading to a very
small $\pi N$ isoscalar scattering length \cite{bkmppn}.
If one moves away from the threshold or the on-shell $\pi N$ kinematics,
however, this cancelation gets
less and less effective. 
Thus the numerical value of the isoscalar re-scattering term
is very sensitive to the details of the individual terms.
Note that, because of
the Weinberg-Tomozawa term,
the proper choice for $q_0$ is also relevant for the
isovector re-scattering that contributes to charged pion production.

If one simply evaluates the re-scattering diagram 
at threshold, neglecting initial- and
final-state interactions,  it is clear
that $q_0=m_\pi/2$.
Keeping this value fixed also when including the distortions 	
leads to an
amplitude that is opposite in sign to the on-shell scattering amplitude,
and interferes destructively with the single-nucleon emission
term \cite{cohen,tree}. This choice for $q_0$ in combination
with the use of the Klein--Gordon propagator for the pion will
be called {\it fixed kinematics approximation} in what follows. 
Refs. \cite{cohen,tree} found that the computed cross sections fell
 well below the data, unless many
other even less-well constrained  terms are included \cite{vkmr}. 
However, once the nucleons are no longer on shell, 
there are other prescriptions in the literature for            choosing
     $q_0$.
In a Feynman diagram this
is the difference between the $0$th component of the nucleon four-momentum
before and after pion emission. Thus one might find it natural to set $q_0$
equal to 
this difference in energies. 
Using this      energy difference prescription
in the  distorted wave Born approximation 
(DWBA) calculation of pion production leads
to a  re-scattering diagram which also
has a sign opposite to that of the single nucleon term, but which
    is about three
times larger in magnitude\cite{sato}.  
As a result, one can reproduce the magnitude of the total cross section
using only the re-scattering diagram. 
This energy prescription will be
called {\it $(E-E')$ approximation} below. 

In addition, having the toy model at hand, we also want to
 study the importance of terms that go beyond the
DWBA, namely the so-called stretched boxes (c.f. Fig.\ref{beitraege}, 
diagrams F3 and F4). These necessarily  occur in
the three-dimensional framework and represent diagrams where there is no 
two nucleon cut.

Note that the questions under investigation
affect not only chiral perturbation theory calculations,
but also more phenomenological approaches.
For example,
Ref.~\cite{teresa} used
the $(E-E')$ prescription for the pion re-scattering when investigating
the influence of nucleon resonances on the production process.
In the so-called J\"ulich model \cite{Han1} 
the full TOPT propagator was used, but 
with its energy fixed to the production threshold.
Thus, a clarification of these formal issues 
is necessary before one 
can draw conclusions about the physics of
the process. This paper is meant to be a step in that direction.

It is important to realize that  one can not resolve the ambiguity 
in the choice of $q_0$ or the proper $\pi NN$ propagator (in what follows
this quantity will sometimes in a somewhat sloppy way be called 
``pion propagator")
by appealing to data. These are questions about the theory which arise
due to the manner 
the DWBA procedure was implemented \cite{cohen,tree,sato,loops}.
 Furthermore, the slow convergence of the momentum expansion requires
one to resolve these difficulties  
before  evaluating  loop diagrams.

  One needs to construct an {\it ab initio} theory of pion production.
Doing this for the realistic case requires that one considers several important
features including 
i) the spin and isospin of
the two-nucleon system;
ii) the Goldstone boson nature of the pion as an odd parity system degenerate
with the vacuum; and
iii) a realistic $NN$ potential.
However, none of these features affects directly the questions
which we want to examine.
Therefore it is appropriate to construct a toy model which is simple enough
to evaluate so that exact answers can be obtained. Then we can consider the
various choices for $q_0$ and for the ``pion propagators" 
as testable approximations.  
In the section II we formulate our toy model,
and examine the various approximations
for final- and initial-state interactions in sections III and IV, 
respectively. Our conclusions are summarized in section V.

\section{The Toy Model}

The first step is to construct the necessary solvable model. Therefore:

i) We consider the production of 
           a scalar ``pion" field which 
has              a Yukawa coupling with 
the nucleons. 
(We shall leave out the quotes around pion in the following text.)

ii) We include two nucleon fields, or,
alternatively, treat nucleons as distinguishable. 
As a consequence,
we need only include pion emission from one nucleon, but not the symmetric
term where the pion is emitted from the other nucleon. 
We do not have to worry about several 
spin-isospin channels, and respective projections.
The simplicity of the model is retained by allowing the pion to
couple to only one nucleon field. As a result, 
              the effects of pion exchange between
two nucleons does not enter. 

iii) A focus of the paper is the pion re-scattering by  one nucleon. 
This pion re-scattering is described by a $\pi N$ seagull vertex 
which is inspired by the
chiral $\pi N$ interaction Lagrangian. 

iv) In order to mock up the nuclear interactions we include 
the exchange of a scalar sigma field, which also       couples to nucleons
via Yukawa coupling. Since the magnitude of this coupling 
has nothing to do with the way to treat the pion energy,
we consider the case of small coupling, and therefore need to only consider
one sigma exchange. 

v) Because $p_i/M_N=\sqrt{m_\pi/M_N}<1$, 
it is typical to treat this problem using
a non-relativistic expansion.
In the following we will examine 
only the leading
terms in this expansion.
In particular, 
contributions from anti-nucleons are not considered.

Therefore, we consider the following toy model defined by the
Lagrangian:

\begin{eqnarray}
{\cal L} = \sum_{i=1,2}N_i^\dagger ( i\partial_0 + \frac{\nabla^2}{2M_N})N_i
+ \frac{1}{2}\left[( \partial_\mu \pi )^2 - m_\pi^2 \pi^2 
+ ( \partial_\mu \sigma )^2 - m_\sigma^2 \sigma^2 \right]
\cr
+ \frac{g_\pi}{f_\pi} N_2^\dagger N_2 \pi
+ g_\sigma \sum_{i=1,2}N_i^\dagger N_i \sigma 
+ \frac{c}{f_\pi^2} \sum_{i=1,2}N_i^\dagger N_i (\partial_0 \pi)^2
\end{eqnarray}
Here $M_N$ is chosen as the physical nucleon mass of 939 MeV and similarly
$m_\pi$ is taken as 139 MeV. 
The mass of the $\sigma$
meson and the cutoff $\Lambda$ on the momentum integrals
are taken as parameters in the theory, to be specified below. 

It is important to immediately display some of the non-realistic 
features of this toy model. 
For simplicity, we did not enforce chiral symmetry,
which would have required a derivative coupling of the pion to 
nucleon spin, instead of the simpler Yukawa coupling.
We are concerned with near-threshold kinematics so that 
a scalar particle 
is  produced  in an  $S$ wave, as is the final $NN$ pair. 
Angular momentum conservation requires  that
the initial  $NN$ pair  also be in an $S$ wave.
In the real world,
however, pions are pseudoscalar and thus the production of $S$-wave pions
calls for a $P$ wave in the initial state. 
Furthermore, 
the toy model includes no strong short-range repulsive $NN$ interactions which
keep the nucleons apart. Thus the nucleons have stronger overlap for our
toy model than in a more realistic treatment. 
However, to a given order in the coupling constants 
we can obtain exact amplitudes for this model,
and are therefore able  to study the
various  treatments of $q_0$ and the $\pi NN$ propagator to determine which, if
any, reproduce the exact model answers.

In a DWBA calculation of threshold pion production, the tree-level
re-scattering diagram 
is influenced 
         substantially by the contributions from the initial and
final state interactions. In this toy model calculation we will therefore 
for simplicity 
concentrate on the DWBA terms where we have only initial or final state $NN$
interactions.  We will in this paper ignore the re-scattering diagram 
with DWBA contributions in 
both initial and final $NN$ interactions 
 since this is a two--loop integration term. Again for simplicity we
will, as discussed, simulate the $NN$ interactions with a 
single $\sigma$ exchange between the nucleons  which occurs before
or after the pion re-scattering process ---the initial-state interaction
and the final-state interaction, respectively. We will discuss these
two cases separately below.
In addition there are graphs in which a $\sigma$ is exchanged in between the
emission and re-scattering of the virtual pion. 
We ignore these here, as they are
not relevant for the issue at hand.
All of our
diagrams are evaluated at order 
$\frac{g_\pi}{f_\pi} g_\sigma^2 \frac{c}{f_\pi^2}$. 
In the following we do not display these  factors, 
as well as other constants which are common to all of the amplitudes.

\section{Final--State Interaction}

\begin{figure}[t]
\begin{center}
\epsfig{file=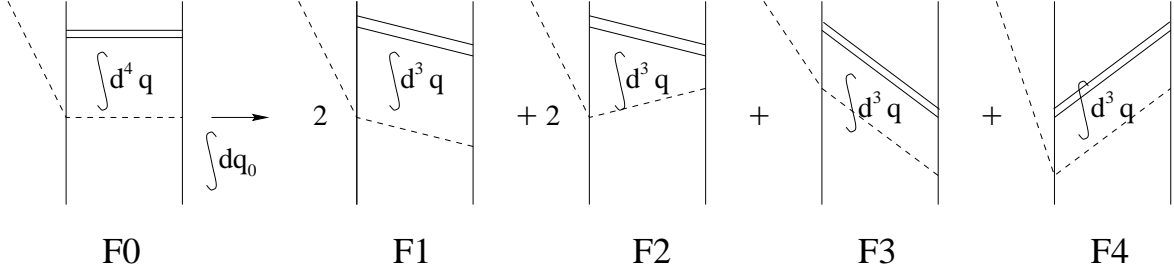, height=3.5cm}
\vskip .25cm
\caption{\it{The diagrams that occur when the sigma exchange appears as
final-state interaction. The analog diagrams I1--I4 with the sigma exchange
in the initial state are considered as well. For the first two diagrams
the two possible time orderings for the sigma exchange lead to
identical expressions.}}
\label{beitraege} 
\end{center}
\end{figure}

The exchange of a $\sigma$ meson in the final state is given 
by the Feynman graph F0 in Fig. \ref{beitraege}. We consider 
threshold kinematics in the center-of-mass frame,
and use the following notation.    
$E$($E'$) represents
the energy of a nucleon in the initial (final) state, 
with  $E_{tot} = 2E = 2E'+m_\pi = m_\pi$ (at threshold: $E'=0$).
In addition, 
$\omega_q = \sqrt{m_\pi^2+\vec q{}\,^2}$ and $\omega_\sigma
 = \sqrt{m_\sigma^2+\vec k{}^2}$ denote
 the $\pi$ and $\sigma$ meson on--shell energies, and 
 $E"=\vec k{}^2/2M_N$ the energy of an
intermediate nucleon. Here $\vec{k} = \vec{p} + \vec{q}$, 
where $\vec{p}$ is
the initial nucleon three-momentum. We choose 
 the pion momentum $q$ to be  the  integration 
variable, so that 
the diagram shown in Fig. \ref{beitraege} (F0) corresponds to the following
four--dimensional integral:
\begin{eqnarray}
\nonumber
\int \frac{d^4q}{(2\pi)^4}  q_0  
& &\left\{ \frac{1}{(E+q_0-m_\pi-E"+i\epsilon)
                                      (E"+q_0-E-i\epsilon)}\right. \\ \nonumber
&\times& \frac{1}{
                                      (q_0-\omega_q+i\epsilon)
                                      (q_0+\omega_q-i\epsilon)} \\
&\times&  \left. \frac{1}{                                                 
                                      (q_0-E+E'+\omega_\sigma-i\epsilon)
                                      (q_0-E+E'-\omega_\sigma+i\epsilon)}
\right\} \ .
\label{fourdim}
\end{eqnarray}

All DWBA calculations are made using a formalism in which
matrix elements are given as three-dimensional integrals. Thus the
first step
is to find the appropriate three-dimensional expression by performing
the $q_0$ integration. Obviously, Eq. (\ref{fourdim}) contains  three poles
in the upper half plane as well as three in the lower half plane. 
One way to proceed would be to close the contour on one of
the half planes and pick each of the three poles enclosed.
However, 
it is more convenient to perform a partial decomposition, in which 
                                                  the poles
 of the pion propagator are isolated before the $q_0$ 
integration is carried out.  It should
be emphasized, however, that the final result does not depend on
the method of its evaluation. 
It should not come as a surprise that the final result of the
$q_0$ integration agrees exactly to the one of TOPT,
as the equivalence between the Feynman prescription
and TOPT is well known.
This is illustrated in Fig. \ref{beitraege},
and the resulting amplitude is given by 
\begin{eqnarray}
\nonumber
\int \frac{d^3 q}{(2\pi)^3}\frac{\omega_q}{4 \omega_q 
\omega_\sigma} & & \left\{ \frac{2}{(E_{tot}-E'-m_\pi -E"-\omega_\sigma)
(E_{tot}-m_\pi -2E")
(E_{tot}-E-E"-\omega_q)}\right. \\ \nonumber
&-& \frac{2}{(E_{tot}-E'-m_\pi -E"-\omega_\sigma)(E_{tot}-m_\pi -2E")
(E_{tot}-E-E"-m_\pi -\omega_q)} \\ \nonumber
&+&\frac{1}{(E_{tot}-E'-m_\pi -E"-\omega_\sigma)(E_{tot}-E'-E-\omega_\sigma
-\omega_q)
(E_{tot}-E-E"-\omega_q)} \\ \nonumber
&-&\left. \frac{1}{(E_{tot}-E'-m_\pi -E"-\omega_\sigma)
(E_{tot}-E-E'-m_\pi -\omega_q-\omega_\sigma)(E_{tot}-E-E"-m_\pi -\omega_q)} 
\right\} \ , \\ 
\label{eq:TOPT}
\end{eqnarray} 
in which 
      the successive four terms can be immediately matched to the diagrams
F1-F4. In particular,  the last two terms are  those of the
stretched box diagrams which have not yet been considered in any 
calculation for pion production. 
We will examine their importance below.
Note, that there is no freedom 
with
respect to what is the appropriate choice for $q_0$ in the numerator of 
Eq. (\ref{fourdim}). 
 The pole structure of Eq. (\ref{fourdim}) in combination
with the way the partial decomposition was performed forces 
$q_0$ = $\omega_q$ in Eq. (\ref{eq:TOPT}), which is an exact equation.

To compare Eq. (\ref{eq:TOPT}) to expressions used in the literature 
it is useful to combine the first two lines to obtain the final state
interaction contribution to the DWBA amplitude:
\begin{equation}
\int \frac{d^3 q}{(2\pi)^3} 
V_\sigma \left(\frac{1}{E_{tot}-m_\pi -2E"}
\right) \frac{m_\pi }{2}G_\pi^{TOPT} \ ,
\label{dwba1}
\end{equation}
where the sigma potential is
\begin{equation}
V_\sigma (k^2)= 
         \frac{1}{\omega_\sigma(E_{tot}-E'-m_\pi- E" -\omega_\sigma)} ,  
\end{equation}
and the TOPT $\pi NN$ propagator ---the exact propagator--- is given by
\begin{equation}
 G_\pi^{TOPT}= \frac{1}{\left(\frac{m_\pi }{2}\right)^2-
                  \left(\omega_q+\frac{\vec{k}^2}{2M_N}\right)^2} \ .
\label{toptpropf}
\end{equation}

Apart from the $\vec{k}^2$ term
in the TOPT $\pi NN$ propagator, Eq. (\ref{dwba1}) 
agrees with  what is known as {\it fixed kinematics approximation}
\cite{cohen,tree}.  
As was explained above, this approximation 
is defined by the use of $m_\pi/2$ for the pion energy in
  both in 
the $\pi N$ seagull vertex  
and in the pionic Klein-Gordon propagator.
In the realistic case (when appropriate nucleon wave functions
are used for the distortions) the 
significance of the $\vec{k}^2/2M_N$  
term in the pion propagator of Eq. (\ref{dwba1}) can be estimated by noting 
that $\vec{k}^2/2M_N$ is of the order of the
the off-shellness of the intermediate nucleons.  Since the final state is
at rest we can estimate 
$\vec{k}^2 = {\cal O}( m_\pi ^2)$ \cite{pwave}.
It then follows in the absence of initial-state interactions
that the loop three-momentum is $|\vec{q}|\sim p_i$.
We can expand Eq. (\ref{toptpropf}) in powers of $m_\pi /M_N$,
and get
\begin{equation}
G_\pi^{TOPT} = G_\pi^{KG}
\left\{ 1 - {\cal{O}}\left(\left(\frac{m_\pi}{M_N}\right)^{\frac{3}{2}}\right) \right\} \ .
\label{approx}
\end{equation}
Here the Klein-Gordon propagator in the fixed kinematics
approximation is defined by
\begin{equation}
G_\pi^{KG}=
\frac{1}{\left(\frac{m_\pi }{2}\right)^2-
 \omega_q^2} \ .
\label{kgprop}
\end{equation}
The 
right-hand side of Eq. (\ref{approx})
 is already expressed in terms of the expansion
parameter of the underlying effective field theory, 
$\sqrt{\frac{m_\pi}{M_N}}$
\cite{cohen,pwave}. Thus ---at the
level of accuracy accessible today--- we expect this Klein-Gordon 
propagator to be a good approximation for
those diagrams where the $NN$ interaction appears in the final state.
Such considerations are not necessarily germane here, however, as
we have not enforced the chiral symmetry on which power counting
is based.
The physical scales appearing in the final state of this model
are set by the parameters $\Lambda$ and
$m_\sigma$, which we take to vary over a large range. 

Let us now  discuss the numerical significance of
the individual terms above. Our toy model allows us to answer the 
following three questions:

\begin{itemize}
\item What is the relative importance of the stretched boxes (F3 and F4 
in Fig. \ref{beitraege}) compared to the ``DWBA--contributions" (F1 and F2)?
\item How good an approximation is the propagator $G^{KG}_\pi$ 
of Eq. (\ref{kgprop}) compared to the exact propagator 
$G^{TOPT}_{\pi}$ of Eq. (\ref{toptpropf})?
\item What is the effect of different treatments of the pion energy ``$q_0$" 
at the $\pi N$ seagull vertex (fixed kinematics compared
to the $(E-E')$ prescription)?
\end{itemize}

\begin{figure}[t!]
\begin{center}
\epsfig{file=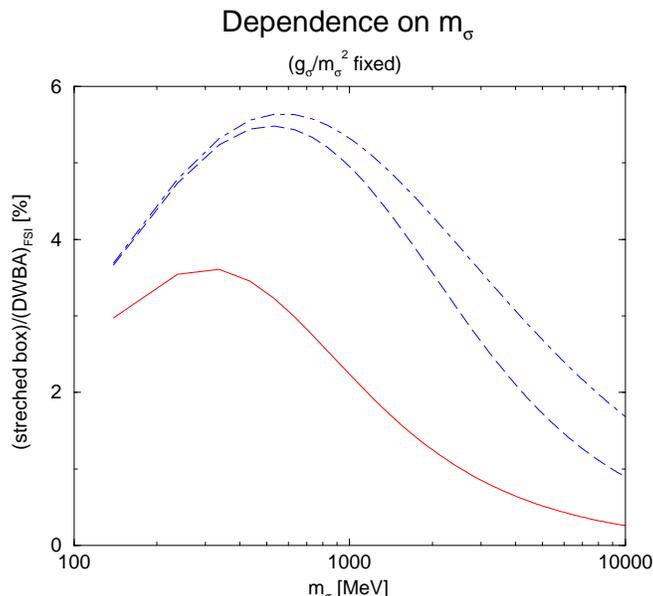, height=8cm}
\caption{\it Importance of the stretched boxes for different choices
of the cutoff as a function of the mass of the $\sigma$ meson.
The ratios of the stretched boxes with respect to the DWBA piece, 
Eq. (\ref{dwba1}), are shown
for $\Lambda=3m_\pi$ (solid line), $10m_\pi$ (dashed line) and $\infty$
(dot--dashed line).}
\label{massdep}
\end{center}
\end{figure}

The answer to the first question is obviously  a function of the 
$\sigma$ mass, since the DWBA contributions should lead to results that
are proportional to $(p_{i}/m_\sigma)^2$, whereas the stretched boxes
lead to $(p_{i}/m_\sigma)^4$. In Fig. \ref{massdep} we show the
ratio of the stretched box contributions to the DWBA part as a function
of the mass of the sigma meson. The three curves correspond 
to three different
values of the cutoff for the radial integration.
As expected, the curves all
fall as $1/m_\sigma^2$ for large $m_\sigma$. 
The 
strength of the stretched boxes never exceeds 6\%. This justifies a 
DWBA treatment of the final state in this  pion production process.

\begin{figure}[t!]
\begin{center}
\epsfig{file=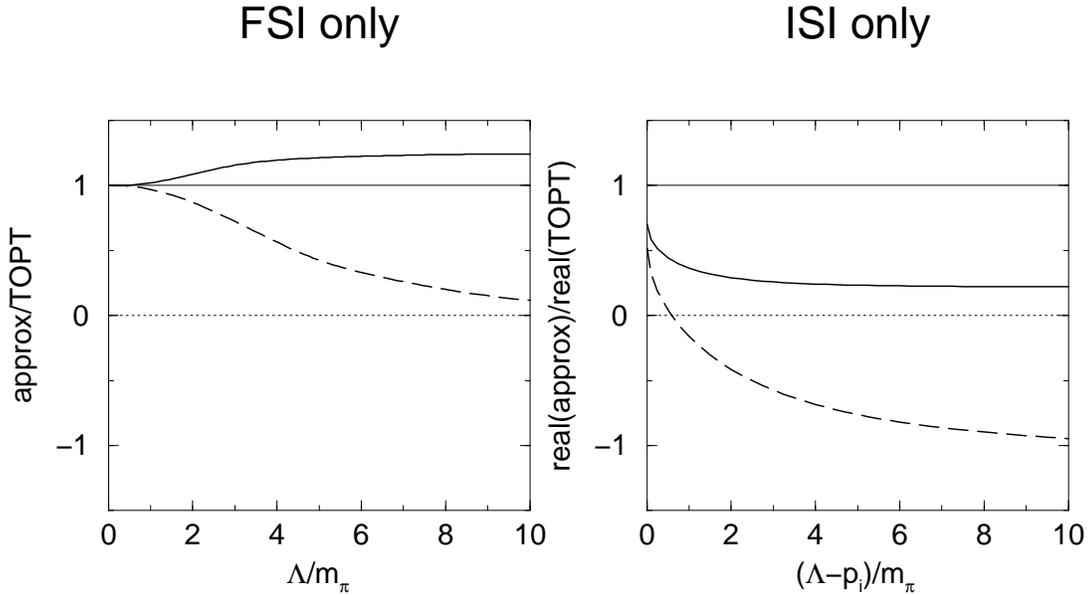, height=8cm}
\caption{\it Effects of the different approximations to the ``production
operator" for different cutoffs. 
The left (right) panel
 shows the result for a sigma exchange in the final (initial)
state in units of the exact answer of Eq. (\protect{\ref{dwba1}})
(Eq. (\protect{\ref{dom}})).
 Using
the ``fixed kinematics approximation" pion propagator 
 leads to the solid curve. 
 The dashed
curve is the result when using the $(E-E')$ approximation. 
  } 
\label{approxtest}
\end{center}
\end{figure}

The answer to the second question is 
presented in the left panel of Fig.~\ref{approxtest} as a function
of the cutoff in the momentum integration. 
We evaluated the 
DWBA piece, Eq. (\ref{dwba1}), with the exact propagator (\ref{toptpropf}) 
and with the approximate propagator (\ref{kgprop}).
The mass of the sigma was chosen to
be $m_\sigma=550$ MeV.
The 
solid curve shows the result using the approximate pion
propagator 
in units of the exact result.
Thus, the
deviation of this approximation from the exact result
never exceeds 25\%.

The third question goes to 
the choice of energy variable ``$q_0$" 
at the $\pi N$ seagull
vertex  reported in the literature \cite{sato}. To simulate this choice we 
replace ``$q_0$" = $m_\pi /2$ in the numerator of Eq.(\ref{dwba1}) with 
\begin{equation}
``q_0" = E-E",
\label{SC}
\end{equation}
 together with the approximate pion Klein-Gordon propagator. 
 This was defined as the $(E-E')$
 prescription above.  In many
reactions Eq. (\ref{SC})  is an appropriate replacement, because
the nucleons remain almost
on the mass shell in the intermediate states. 
However, as soon as large intermediate
momenta are accessible, this treatment might be questionable. 
In Fig. \ref{approxtest} the
dashed curve
shows the results of using the $(E-E')$ prescription, again
in units of the exact result. 
Within our toy model, the result shows that
this approximation is not reasonable for calculations of
 threshold pion production. 
Note that this amplitude  is very sensitive to the sigma mass,
which acts as a regulator.
If the sigma mass is taken to be larger, then the result
changes even more dramatically with the cutoff.
A change in sign happens at the point where the cutoff is
big enough for the effect of the 
$\vec{k}^2/2M$ to overcome that of $m_\pi /2$
(the larger the intermediate
momentum, the larger $E"$). 

The net result of the toy model 
for the final-state interaction case is that
 using $``q_0"=m_\pi/2$ in both the virtual $\pi N$ seagull scattering vertex 
 (numerator) and in the approximate pion Klein-Gordon propagator
 is very reasonable.  

\section{Initial--State Interaction}

We now consider 
the case when the sigma exchange occurs before the re-scattering process. 
A reduction to the three-dimensional
integral (or starting with the TOPT expression) 
gives the following four terms:  
\begin{eqnarray}
\nonumber
M_I  =  \int \frac{d^3 q}{(2\pi)^3}
               \frac{\omega_q}{4 \omega_q \omega_\sigma} & &
\left\{ 
      \frac{2}{(E_{tot}- E' - \bar E" -\omega_q) 
             (E_{tot}- 2  \bar E" ) (E_{tot} - E -  \bar E" - \omega_{\sigma})}
  \right. \\ \nonumber
 & & -\frac{2}{(E_{tot}- E' - \bar E" -\omega_q - m_\pi)
            (E_{tot}- 2  \bar E" ) (E_{tot} - E -  \bar E" - \omega_{\sigma})} 
  \\ \nonumber
  &  &   
 + \frac{1}{(E_{tot}- E' - \bar E" -\omega_q)
 (E_{tot}-E-E'-\omega_q -\omega_{\sigma})(E_{tot}-E-\bar E" -\omega_{\sigma})}
    \\ 
  & &   \left.
  -\frac{1}{(E_{tot}- E' - \bar E" -\omega_q - m_\pi)
(E_{tot}-E-E'-\omega_q - \omega_{\sigma} - m_{\pi})
(E_{tot}- E - \bar E" - \omega_{\sigma})}
 \right\} , 
\label{mi}
\end{eqnarray}
where again
$\omega_q  =  \sqrt{m_\pi^2 + \vec{q}\;^2} \;,
\omega_\sigma  =  \sqrt{m_\sigma^2 + (\vec{q} + \vec{p})^2}$,  
and $E' = 0$,
but the energy of an intermediate nucleon is
$\bar E" =  \vec{q}\;^2/2M_N$.

As before, the first two terms in Eq. (\ref{mi}) correspond to box diagrams
and the last two to stretched boxes, cf. Fig. \ref{beitraege}.
In case of the initial-state interaction 
the stretched boxes 
still turn
out to be 
smaller than the boxes, but less so:
$\simeq 30 \%$.
Note that
this is also of the size expected in the real world
where the
expansion parameter of the EFT is 
$\sqrt{\frac{m_\pi}{M_N}}
\simeq 0.4$
\cite{cohen,pwave}.
We therefore concentrate on those terms containing
the $NN$ propagator, $G_{NN} = (E_{tot} - 2 \bar E" )^{-1}$, only
and obtain
\begin{eqnarray}
\int \frac{d^3 q}{(2\pi)^3} \frac{m_\pi  }{2}\; G_{\pi NN}^{TOPT}
 \left(\frac{1}{E_{tot}-2\bar E"}\right)
V_\sigma ,
\label{dom}
\end{eqnarray} 
where  
\begin{eqnarray} 
V_\sigma (\vec{q}\;^2,\vec{k}\;^2)= 
\frac{1}{\omega_\sigma(m_\pi/2-\bar E" -\omega_\sigma)} , 
\end{eqnarray}
and in the initial-state interaction
case the TOPT $\pi NN$ propagator reads
\begin{equation}
G_{\pi NN}^{TOPT} =
\frac{1}{\left(\frac{m_\pi }{2}\right)^2-
                  \left(\omega_q+\bar E" -\frac{m_\pi }{2}
                               \right)^2}
\ . 
\label{toptprop}
\end{equation}
This looks like a DWBA expression using the $m_\pi/2$ prescription.
Due to the large initial momentum, the unitarity cut of  
$G_{NN}$ turns out to be an essential feature.

Similar to the section on final-state interaction, we investigate the 
{\it fixed kinematics
approximation} and the $E-E'$ {\it approximation} using the free 
pionic Klein-Gordon 
propagator, Eq. (\ref{kgprop}). This means especially
that in the $G_{\pi NN}^{TOPT}$ 
of Eq.(\ref{toptprop}) we set $\bar E"$ = $m_\pi /2$, which implies 
on-shell intermediate nucleons: $\vec{q}\,^2 = m_\pi M_N$. 
In the $E-E'$ {\it approximation} we further replace $``q_0"=m_\pi/2$
by $E-\bar E"$.
In the right panel of Fig. \ref{approxtest} we demonstrate 
the inadequacy of both approximations,
compared to the exact result given by Eq. (\ref{dom}). 

Due to the large initial momentum the imaginary
part of these diagrams turns out to be of the order of  the real part.
(Since we work at the kinematical threshold of pion production the 
imaginary part
from $G_{\pi NN}$ is zero). Since all the approximations where constructed
such that they agree once the intermediate two-nucleon state goes on-shell,
all the individual results agree for the imaginary part.

The question becomes why do both
 approximations show such a large deviation from
the exact result of Eq. (\ref{dom}). 
The cause can be traced back to the appearance of a $\pi NN$  cut in
the exact propagator:
from Eq. (\ref{toptprop}) 
we see that the propagator $G_{\pi NN}^{TOPT}$
diverges as $|\vec{q}|^{-2}$ when $\vec{q}$ approaches 0.
On the other hand
 for small $\vec{q}$ the free pionic Klein-Gordon 
propagator $G_\pi^{KG}$ goes to a constant. 
It is the very different nature of the infrared behaviors of the propagators,
$G_\pi^{KG}$ and $G_{\pi NN}^{TOPT}$, that leads to the large deviation
of (the real part of) the amplitude 
from the result of 
Eq. (\ref{dom}).

Having identified the $\pi NN$ cut as an important feature of the
production reaction, a natural question that arises is how to
set up a counting scheme capable of covering this. Note that contrary
to the more conventional contributions, where the scale of typical
momenta is set by the initial momentum $p_i = \sqrt{M_N m_\pi}$, the
$\pi NN$ cut pronounces momenta of the order of the external pion
momentum. It can not be a part of a toy-model investigation to completely
resolve this matter ---after all our model interaction is not consistent
with the requirements of chiral symmetry. However, we will use the
last part of this section to suggest a possible method to address
the issue.

To this end we will 
rewrite Eq. (\ref{mi}) such that we isolate both the $NN$ and the $\pi NN$ 
singularities. For this purpose we are guided by 
the unitarity transformation method of 
Ref. \cite{unitary,unitary2}. 
This method is one way to isolate
the different singularities of a particular diagram.
In this case
the scattering amplitude can be written as
(where for clarity we suppress $\int d^3q $ as well as some over all factors)
\cite{poles}: 

\begin{eqnarray}
M_I & = & M_{NN} + M_{\pi NN} + \cdots , \label{set1} \\
M_{NN} & = & \frac{1}{m_\pi-2\bar E"}
 \left(\frac{m_\pi /2}{\bar E"^2 - \omega_q^2}\right)
  V_\sigma , \label{NN2} \\
M_{\pi NN} & = & - \frac{1}{m_\pi-\bar E"-\omega_q}
  \left(\frac{m_\pi /2}{\bar E"^2 - \omega_q^2}
  \right)    V_\sigma ,
  \label{pin1} 
\end{eqnarray}
where the ellipsis denote the stretched box TOPT
diagram contributions. 
We see that 
the above amplitude has two physical singularities due to the $NN$ and $\pi NN$
scattering states. 
We find numerically that
$M_{\pi NN}$ is about 5 times larger than $M_{NN}$ when
evaluated with a cutoff of $\Lambda$ = 10$m_\pi$.
This large effect of the $\pi NN$ cut in the toy model is also responsible for
the stronger effect of initial-state stretched boxes,
as the latter
also contains the $\pi NN$ cut (see fourth line
of Eq. (\ref{mi})).
These two points highlight the numerical significance of the 
three-particle cut. 

Note that
this is not a unique separation of the two branch cuts. 
To make closer contact with previous work \cite{unitary,unitary2}
we can rewrite Eq. (\ref{set1}) in a form closer to the
$(E-E')$ {\it prescription}:
\begin{eqnarray}
M_I & = & M'_{NN} + M'_{\pi NN} + \cdots ,\label{set2} \\
M'_{NN} & = & \frac{1}{m_\pi-2\bar E"}
 \left(\frac{\bar E"}{\bar E"^2 - \omega_q^2}\right)
  V_\sigma ,  \label{NN'2} \\
M'_{\pi NN} & = & - \frac{1}{m_\pi-\bar E"-\omega_q}
  \left(\frac{(\bar E"+\omega_q)/2}{\bar E"^2 - \omega_q^2}
  \right)    V_\sigma . 
  \label{pin2} 
\end{eqnarray}
Clearly $M_{NN} + M_{\pi NN}$ = $M_{NN}' + M_{\pi NN}'$,
although some shift of strength is then achieved between 
$NN$ and $\pi NN$ contributions: in this case
we find the contribution from $M_{\pi NN}'$ larger in magnitude
than $M_{NN}'$ by a factor 2, using the same cutoff $\Lambda$ = 10$m_\pi$.
It remains to be seen
which splitting is the most appropriate in the realistic case.

The most significant finding for the case of the initial-state interaction
is therefore that in the toy model
the three-body $\pi NN$ branch cut of $G_{\pi NN}$ is very important. 
The importance of this cut has been advocated before,
for example in Ref.\cite{blankleider}. 
Here the static propagator, which was defined as being part of the {\it fixed
kinematics approximation} as well as of the 
$(E-E')$ {\it approximation}, leads to
erroneous results for the real part of the amplitude. 

However, it is important to remark that we expect the importance
of this branch cut to be much smaller in the real world.
Indeed, as we have seen, close to threshold this 
type of contribution comes from
three-momenta near 0. In the real world, chiral
symmetry suppresses such contributions.
The pion coupling in leading order in chiral perturbation theory,
for example, goes through the pion three-momentum.
In our toy, chiral
symmetry does not play a role, the pion coupling 
is a simple Yukawa coupling, and
both initial and final $NN$ 
states are in relative $S$ waves, which enhances the influence
of the $\pi NN$ cut.
The power counting developed in Refs. \cite{cohen,pwave} does
take into account
chiral symmetry ---thus the correct
factors of momenta--- and suggests a suppression of 
these branch effects, as long as the momentum of the emitted
pion is ${\cal O}(m_\pi)$ (or less).
Clearly, it is important
to further study the power counting 
---in particular in conjunction with the unitary 
transformation method--- in the realistic case. 

\section{Summary and Conclusion}

We have investigated various approximations for pion production by defining
a toy model which allows the computation of exact model 
transition-matrix elements.
Because it lacks chiral
symmetry, this model has the unrealistic 
features that both the initial and final state
$NN$ wave functions are $S$ states. 
The influence of
$NN$ correlations that suppress the short-distance wave functions 
are absent from the toy model.
Furthermore, diagrams with both initial- and final-state interactions
could also be important in more realistic calculations.
We have performed
some test calculations using  the Reid $NN$ potential,
which indicate 
that $NN$ correlations do modify
some of the toy model findings at a quantitative level. However, the 
toy model allows the compilation of exact results at a given
order in the couplings, and thus some qualitative insight.

The findings of this paper can be summarized as follows:

\begin{itemize}

\item The stretched box contributions are numerically small 
compared with boxes.

\item For the final-state interaction, only the {\it fixed
kinematics approximation} (for both propagator and
vertex) turns out to be appropriate. 

\item If a loop with the initial-state interaction is included, 
the contribution of the $\pi NN$ cut is very important and
has to be taken into account
properly, which is not done in the common approximations.

\end{itemize}

The first two findings
are in accord with the expectation from the existing power counting
for pion production
in the effective field theory \cite{cohen,pwave}.
Indeed, according to this power counting  
stretched boxes 
involving pions are
sub-leading and 
those involving heavier mesons are absorbed in higher-order local
operators.
Moreover, due to infrared enhancements that lead to 
the (quasi) bound state in the $NN$ interaction,
the
effect of the final-state interaction in realistic calculations 
should be by far dominant close to
threshold.

The third finding is perhaps surprising.
However, chiral symmetry is expected to be crucial
in suppressing this contribution in the real world, 
because the $\pi NN$ cut emphasizes small momenta.
Clearly, the importance of the three-body nature of the intermediate state
needs to be further
examined in realistic calculations.

\vskip 1cm

\noindent
{\bf Acknowledgments:} 
We would like to thank Harry Lee for hosting the INT/Argonne
Workshop on Pion Production, where this work originated.
For hospitality,
FM and UvK thank the Nuclear Theory Group and
the Institute for Nuclear Theory at the University of Washington,
and TS thanks the Nuclear Theory Group 
at the University of South Carolina. 
CH would like to thank the Alexander von Humboldt
Foundation for financial support. 
We thank the U.S.
Department of Energy [grant DE-FG03-97ER41014 (GAM and UvK)] and
the NSF [grants PHY-9900756 (FM) and PHY 94-20740 (UvK)]
for partial support.
UvK would also like to thank 
RIKEN, Brookhaven National Laboratory and the U.S.
Department of Energy [grant DE-AC02-98CH10886] for providing the facilities
essential for the completion of this work.


\begin{thebibliography}{99}
\bibitem{poles}Note, in the non-relativistic treatment the individual terms
employ poles at $\bar E"\simeq 2M_N$. However, since this kinematic regime 
is beyond the applicability of the non-relativistic treatment anyhow and
the cutoff is chosen to be well below this scale, we do not need to worry
about these in the context of this presentation.
\end{thebibliography}
\end{document}